\documentclass[english,12pt]{article}
\usepackage[twoside,a4paper]{geometry}
\usepackage[T1]{fontenc}
\usepackage[latin9]{inputenc}
\usepackage{amsthm}
\usepackage{amsmath}
\usepackage{amsfonts}
\usepackage{babel}

\begin{document}
\title{\Large On the history of Levi-Civita's parallel transport}
\author{\small Giuseppe Iurato\\\small University of Palermo, Palermo, IT\\\small E-mail: giuseppe.iurato@community.unipa.it}
\date{}
\maketitle
\thispagestyle{empty}
\begin{abstract}\noindent
In this historical note, we wish to highlight the crucial conceptual role played by the principle of virtual
work of analytical mechanics, in working out the fundamental notion of parallel transport on a Riemannian
manifold, which opened the way to the theory of connections and gauge theories. Moreover, after a detailed
historical-technical reconstruction of the original Levi-Civita's argument, a further historiographical
deepening and a related critical discussion of the question, are pursued\textsuperscript{,}\textsuperscript{,}\footnote{{\it 2008 PACS}: 02.40.Ky, 02.40.Yy, 01.65.+g, 45.20.Jj,
11.15.-q.}.\end{abstract}

\section{Introduction}
The role of the principle of virtual works of analytical mechanics, in formulating Levi-Civita's parallel
transport of Riemannian geometry, has already been emphasized in \cite{ir}, to which we refer for the first
prolegomena to the question as well as for a wider and comprehensive historical contextualization of it. In this
note, which falls into the intersection area among history of mechanics, history of differential geometry and
history of theoretical physics, we would like to deepen this aspect regarding the genesis of one of the most
important ideas of modern mathematics and its applications to theoretical physics (above all, field theory),
enlarging the framework of investigation with the introduction and critical discussion of further, new
historiographical elements of the question, after having reconstructed technically the original argument of
Levi-Civita. In what follows, therefore, we briefly recall, as a minimal historical introduction, some crucial
biographical moments of the Levi-Civita, which will help us to better lay out the question here treated.

To be concise, Tullio Levi-Civita, besides to have been one of the most important mathematicians of 20th century
with wide interests in physics and mathematical physics (cf. \cite{leci3}), was, in particular, a clever master
in applying methods of absolute differential calculus to other subject-matters which go beyond pure mathematics,
above all mechanics and mathematical physics, and vice versa, as emerges, for instance, from a detailed
historical enquiry of the seminal paper \cite{leci2} -- whose first results have been exposed in \cite{ir} --
which was conceived, by the Levi-Civita, in a well-determined moment of his academic and scientific career, that
we rapidly sketch out as follows.

Soon after he graduated at Padua in 1892 with Gregorio Ricci-Curbastro, Levi-Civita was appointed, in 1895, as
internal professor to the high school attached to the Faculty of Science of Pavia University. In this period,
Levi-Civita started his academic career and research activity with some notable studies in analytical mechanics
and higher mechanics, where, for the first time, he applied the new methods of absolute differential calculus to
approach and solve certain open problems of mathematical physics. In 1896, Levi-Civita moved to Padua, to the
chair of rational mechanics (left uncovered by death of Ernesto Padova), who held for more than two decades. In
this place, comforted by the quite life of his own home town, he continued with research in mathematical
physics, with particular attention to higher mechanics and its applications. Just in this period, Levi-Civita
conceived his famous work \cite{leci2}, here in historical quest.

The Levi-Civita's research in Padua period, was fully devoted to mathematical physics arguments, with some
occasional work on pure mathematics but ever motivated to answering to some technical issues inherent formal
methodologies of treatment of mathematical physics questions and problems. However, this line of research was
yet motivated by the cold reception of the new methods of absolute differential calculus by international
mathematical community, a formal system conceived just by Ricci-Curbastro and Levi-Civita. In fact, last
Levi-Civita's work published on this argument, was the well-known report on absolute differential calculus,
required by Felix Klein for the {\it Mathematische Annalen}, at the turn of 19th century (cf. \cite{rl}), and
written with his schoolmaster Ricci-Curbastro.

It was only after absolute differential calculus turned out to be at the formal basis of general relativity
theory that Levi-Civita willingly turned back to this argument, around the late of 1910s, starting just with the
seminal paper\footnote{In the follows, when we quote \cite{leci2}, we always refer to the version contained in
the Volume 4 of the collected papers of Levi-Civita \cite{leci3}.} \cite{leci2}, with which he opened as well
new perspectives in the mathematical treatment of relativistic theories after having had a crucial
correspondence just with Albert Einstein, at the beginnings of 1915, on some crucial, yet problematic, formal
aspects of his well-known gravitational field equations, which Levi-Civita definitively clarified. In 1919,
Levi-Civita moved to Roma, first to the chair of higher calculus, whose lecture notes, centred on absolute
differential calculus, were recollected in \cite{leci0}, then to the chair of rational mechanics. The decennial
experience in teaching the latter subject-matter, was concretized in the celebrated treatise \cite{leci},
written in collaboration with Ugo Amaldi, whose first edition dates back to 1923.

In regard to the seminal paper\footnote{Strangely enough, this work seems have not been translated in any other
foreign language.} \cite{leci2}, our main intention, here, is above all to show how a basic formal tool of
analytical mechanics, like the principle of virtual works in the Lagrangian formulation, was ingeniously used
first to sketch, then to develop formally, the geometric idea of parallelism on a Riemannian manifold, as
originally conceived by Levi-Civita. With respect to what has been said in \cite{ir}, in this note we would like
technically to deepen just this founding moment of the pathway followed by Levi-Civita in developing his
celebrated and fruitful idea, as well as to add further historiographical data with a related critical
discussion and examination. What will emerge from that, will be the extreme simplicity and elegant formal style,
together the high intuitive charge and clarifying powerfulness, of this typical method of study and research
performed by Levi-Civita in almost all his works.

Nevertheless, just in regard to \cite{leci2}, in most of the related mathematical literature\footnote{Of which
we quote \cite{stru}, p. 4; \cite{kob}, Vol. I, p. 287; \cite{maur}, p. 19. In particular, we also cite
\cite{sha}, App. A, Sect. 1, p. 358, which even says textually that: {\it ''Perhaps the motivating reason for
Levi-Civita's search for what is now called the Levi-Civita connection was the need to find an analog of the
vector gradient of Euclidean geometry {\rm [...]} in the context of Riemannian geometry. In fact, Levi-Civita
called his discovery the {\rm absolute derivative}''}.}, we often find the statement according to which
Levi-Civita's parallel transport was motivated by the attempt to give a geometrical interpretation to the
so-called {\it covariant derivative} of absolute differential calculus, a tool formally explained in \cite{rl}
(on the basis of previous ideas of Elwin B. Christoffel), which was indeed provided later not by Levi-Civita,
but rather by others (among whom are Hermann Weyl and Gerhard Hessenberg). Instead, if we read carefully his
paper \cite{leci2}, one sees that Levi-Civita wasn't never motivated by this end, but rather by the will to
simplify the computation of the curvature of a Riemannian manifold, re-examining the covariant behavior of
Riemann's symbols and their occurrence in relativistic questions (cf. \cite{leci2}, Introduction).

On the other hand, even if Levi-Civita was aimed by the attempt to enquire on covariant behavior of those
geometric entities involved in the computation of Riemannian curvature, mainly Riemann's symbols, the covariant
derivative was not yet the predominant, central formal tool of absolute differential calculus involved in his
investigations which, as we shall see later, were merely carried out within a geometrical framework set up
according to an analytical mechanics standpoint, as an intuitive guide for the next analytical calculation of
curvature, independently of the covariant derivative algorithm. This conclusion may be also inferred by what
Levi-Civita himself said at page 7 of the preface to \cite{leci0} (to which we refer for more details), namely,
that the close relationships between covariant derivative and parallelism were explicited later by others
scholars, not by him. But, just to consider the numerous, new results and improvements coming from his notion of
parallelism, Levi-Civita held two courses at Roma University in the years 1922-23, whose lecture notes were
collected and drawn up by Enrico Persico in \cite{leci0}.

As Levi-Civita himself said in the Introduction to \cite{leci2}, initially he was motivated by the intention to
simplify the methods of calculus of the curvature of a generic Riemannian manifold, involving Riemann's
symbols\footnote{See also the obituary by Carlo Somigliana \cite{som}, who has rightly understood historically
which Levi-Civita's real intentions led him to draw up \cite{leci2}. Somigliana also reminds a later, more
intuitive interpretation of this notion of parallelism over an hypersurface of $S_N$, given, by Levi-Civita
himself, in \cite{leci0}, as envelope of tangent spaces along a certain curve lying on that hypersurface. This
is the same intuitive (geometrical) interpretation given by \cite{col}, Ch. V, Sect. 1, arising from some rigid
body kinematics interpretations of Levi-Civita's parallel displacement given later by many authors (among whom
are E. Persico and G. Corbellini -- cf. \cite{leci0}, Ch. V-(b), pp. 119 and 121).}. To this end, a preliminary
geometric sight of this question, allowed him to work out an idea of parallelism on a Riemannian manifold which
turned out to be a needful notion for the computation of the curvature of this manifold according to the usual
formal methods of the time, basically centred on the possible vectorial circuitations along suitable
infinitesimal closed contours (among which are the so-called {\it geodesic parallelogrammoids}) lying on the
given manifold, hence considering the related commutation properties of certain geometrical parameters or
entities closely related with the circuitating vectors\footnote{As, for instance, the independent infinitesimal
displacements $dx_i$ and $\delta x_j$, involved in \cite{leci2}, Sect. 15., and representing generic tangent
directions to certain geodesic curve traits. See, above all, \cite{lic}, Ch. V, Sect. III, No. 80; see also
\cite{pal1}.}.

In pursuing this research program for the computation of the curvature of a Riemannian manifold (which will led
to the so-called {\it Levi-Civita's geometric characterization of Riemannian curvature}\footnote{Cf. \cite{ros},
Ch. 8, pp. 316-317.}), Levi-Civita devoted the first fourteen sections of his memoir -- i.e., most of the whole
paper -- just to introduce and explain the new concept of parallelism upon an arbitrary Riemannian manifold
$V_n$ with dimension $n\geq 2$, embedded in some ordinary Euclidean space $\mathbb{R}^N$ (who he denotes with
$S_N$). Therefore, he applied this new geometric idea for simplifying the computation of Riemannian curvature,
hence exposed the achieved results in the sections 15 and 16 of \cite{leci2}. Levi-Civita himself said
textually, in the Introduction to \cite{leci2}, that such a paper originally sprung out only for this end, but
that accordingly it was then enlarged consistently to provide as well the related geometrical interpretation,
not to covariant derivative, but to this new geometric trick for calculating curvature. Likewise, in the
contents of \cite{leci0}, parallelism is exposed (at Ch. V) before covariant derivative (at Ch. VI), not vice
versa.

This latter geometrical interpretation, therefore, as again Levi-Civita pointed out in the Introduction to
\cite{leci2}, was referred to the notion of curvature of a Riemannian manifold, as originally conceived,
although quite implicitly, by Riemann in his celebrated 1854 inaugural lecture {\it \"{U}ber die Hypothesen,
welche der Geometrie zu Grunde liegen}, and not to other, like covariant derivative, this latter having been
very far from the manifest intentions that really pushed Levi-Civita to drawing up his work. He, more times, in
his memoir of 1917, made reference to this Riemann seminal paper, as well as to the related Heinrich Weber {\it
Commentatio mathematica} (Part II) in which further formal remarks and details to Riemann's work of 1854, are
exposed\footnote{Cf. \cite{rie}, XXII.}. So, the main aim which led Levi-Civita to draw up his memoir of 1917,
was essentially to deepen and clarify pioneering Riemann's ideas on the curvature of metric manifolds, as
sketched in the famous inaugural lecture of 1854. Levi-Civita succeeded in this difficult task, in a very
elegant and intuitive fashion, by means of an analogical-conceptual transcription of the initial pure geometry
question into a suitable analytical mechanics framework.

Levi-Civita, afterwards, devoted the last two sections 17 and 18 of \cite{leci2}, just to explicitate what he
deemed were implicitly present in the original Riemann's memoir, just through those suitable geometrical tools,
primarily his notion of parallelism, which showed to be indispensable to accomplish this task, but not expressly
considered by Riemann. In these sections of \cite{leci2}, Levi-Civita exactly used his notion of parallelism on
a generic Riemannian manifold, as applied in the previous sections 15 and 16 of \cite{leci2}, to clarify and
determine easier the covariant behavior of Riemann's symbols as well as the curvature of a Riemannian manifold
with a generic metric, by means of that usual method making reference to infinitesimal geodesic
parallelogrammoids and related commutation properties of the first-order infinitesimal displacement operators
$d$ and $\delta$.

Anyway, what we wish to highlight in the present note, is that, by examining technically the main passages of
his construction of the notion of parallelism, one becomes even more aware that Levi-Civita's strong training in
analytical mechanics surely had a primary, central role in the related conceptual developments of this idea,
influencing so deeply the related logical reasoning fashion in such a way that intuition and insight earned
much; but, at the same time, all that guided Levi-Civita in his analytical and rigorous treatment of the
question, easily reaching to the final local differential equations of parallelism. This may be also due to the
fact that geometry and mechanics then had (and, in a certain sense, still have) evanescent boundaries, wide
intertwinement zones and a common language with very much similar traits and meaning analogies, with mutual
benefit. In what follows, we shall try to reconstruct, explain and justify, in a detailed manner, the crucial
points of the original Levi-Civita's proceeding in which are masterly involved concepts and methods of
analytical mechanics together differential geometry arguments.

\section{A brief recall on the principle of virtual works}
In this section, we spend only a very few words on the principle of virtual works, referring to \cite{ir} for a
deeper historical discussion of it. To be closest to Levi-Civita, as well as for a more methodological coherence
with the historical issue here treated, we shall mainly follow his celebrated treatise on rational mechanics
\cite{leci} to briefly enunciate the principle of virtual works in its formal essence, to turn out enough for
our aims.

The fundamental dynamical equations of a generic system of $N(\geq 1)$ material points of mass $m_i$, subjected
to preassigned {\it active forces} $\vec{F}_i$ and {\it constraint reactions} $\vec{R}_i$, may be written as
follows
\begin{equation}
(\vec{F}_i-m_i\vec{a}_i)+\vec{R}_i=\vec{0},\qquad i=1,2,...,N,\label{equa1}
\end{equation}where $-m_i\vec{a}_i$ are said to be {\it inertial forces}. These equations are the formal statement of the
so-called {\it D'Alembert's principle}, which allows to reduce formally any dynamical problem to a static
one\footnote{Cf. \cite{leci}, Vol. I, Ch. XV, Sec. 1, No. 2; Vol. II, Part I, Ch. V, Sec. 3, Nos. 18-20.}. This
principle, which may be enunciated in many, yet equivalent, fashions, is, together Lagrange's principle of
virtual works (see later), one of the key principles of analytical mechanics.

The {\it principle of virtual works}, in its original Lagrangian formulation, states that the work of the
reactions $\vec{R}_i$, due to smooth constraints, is non-negative for any irreversible virtual displacement,
while is zero for any reversible virtual displacement\footnote{Cf. \cite{leci}, Vol. I, Ch. XV; Vol. II, Part I,
Ch. V, Sect. 3, Nos. 18-21; \cite{leci1}, Vol. I, Ch. XIV, Sect. 2, Nos. 4-8; Vol. II, Ch. V, Sect. 3, Nos.
17-19; \cite{agopi}, Vol. II, Ch. V, Sect. 1, No. 4; \cite{ago}, Ch. I, Sects. 1-2; \cite{fin}, Vol. 1, Ch.
XIII, Sect. 4.}. In the case of bilateral, smooth constraints, as expressed, for example, by $r$ equalities
providing equations of a smooth manifold of codimension $r$, all compatible virtual displacements are
reversible, so we have that the virtual work of constraint reactions is zero, whence the principle of virtual
works reads
\begin{equation}
\delta L=\sum_i \vec{R}_i\cdot\delta\vec{P} _i=0,\label{elv}
\end{equation}
where $\delta\vec{P} _i$ is the first-order virtual displacement of the point of application of $\vec{R} _i$.
Relation \eqref{elv} is also said to be the {\it symbolic equation of statics}\footnote{Also said to be {\it
D'Alembert-Lagrange principle} as reformulated by Lagrange (cf. \cite{arno}, Ch. IV), or {\it general equation
of virtual work} (cf. \cite{bell}, Vol. I, Ch. XV, Sect. 318; \cite{kra}, Vol. I; \cite{somm}). See also the
references quoted in the previous footnote.}.

If we accept, following Lagrange\footnote{Cf. \cite{cape}.}, that inertia is another force, then it should be
add to the active ones. Therefore, from \eqref{equa1} and \eqref{elv}, it follows that Lagrange's principle of
virtual work states that a finite system of material points is balanced when the active forces $\vec{F}_i$, to
which it is subjected, satisfy\footnote{Cf. \cite{gol}, Ch. 1, Sect. 1.4, Eqs. (1.43)-(1.45); \cite{spiv2}, Part
I, Ch. 6, p. 210; Part III, Ch. 12, p. 441.}
\begin{equation}
\delta L=\sum_i\vec{F}_i\cdot\vec{\delta P}_i=0,\label{elvlagr}
\end{equation}
where $\delta\vec{P}_i$ is the first-order infinitesimal displacement of the application point of $\vec{F}_i$.
Thus, if a system is at equilibrium, the virtual work of all active forces $\vec{F}_i$ will vanish for any
virtual displacement. Relation \eqref{elvlagr} is also said to be the {\it symbolic equation of
dynamics}\footnote{Cf. \cite{leci}, Vol. II, Part I, Ch. V, Sect. 3, No. 20.}.

If the constraints are holonomic, then they are expressed as equalities in the intrinsic parameters of the
system\footnote{Cf. \cite{leci}, Vol. I, Ch. VI, Sects. 1 and 3.}, and the vanishing of the virtual work of
constraint reactions assumes an interesting geometrical meaning, as we shall see later. In particular, in the
case of a material point constrained to lie upon a smooth surface (or a smooth curve), then we have a reaction
which is perpendicular to the surface (or to the curve), while every first-order virtual displacement lies on
the tangent plane (or on the tangent line) of the surface (or of the curve). Exactly in this case, the
constraint reaction therefore spends no work\footnote{Cf. \cite{leci}, Vol. I, Ch. XV, Sec. 1, No. 3-a).}.

This latter case-study will be a very emblematic one in the following historical enquiry, when we shall discuss
how and why the symbolic equation of dynamics \eqref{elvlagr}, was so crucial in developing Levi-Civita's notion
of parallel transport on a Riemannian manifold, bringing back, in doing this, to the consideration, following
Levi-Civita, of a mechanical conceptual analogy just with this case-study.

\section{On Levi-Civita's parallel transport}

Our historical method basically consists in a careful reading and in a detailed historical analysis primarily of
the original sources related to the question here under investigation. In our case, therefore, we strictly
follow first the original paper of Levi-Civita, i.e., \cite{leci2}, hence other possible related works of the
same author. In the next section, then, we shall consider, historiographically, part of the secondary literature
on the argument, whose sources, discussed and criticized from an historical standpoint, will allow us to do new,
further remarks and hints on the question.

In \cite{leci2}, Sect. 1, Levi-Civita begins with the consideration of two arbitrary directions
$\vec{\alpha},\vec{ \alpha}'$ emerging from two infinitesimally near points $P,P'$ of a generic Riemannian
manifold $V_n$, embedded in an $N$-dimensional Euclidean space $S_N$ (i.e., $\mathbb{R}^N$) of suitable
dimension $N$. Thinking $V_n$ as immersed into $S_N$, we may start considering $\vec{\alpha}$ and
$\vec{\alpha}'$ in $S_N$, where the Euclidean geometry condition of parallelism implies that these two
directions $\vec{\alpha},\vec{\alpha}'$ are parallel if and only if
\begin{equation}
\mbox{angle}\left(\widehat{\vec{\alpha},\vec{f}}\right)=\mbox{angle}\left(\widehat{\vec{\alpha}',
\vec{f}}\right)\label{para}
\end{equation}
for any auxiliary direction $\vec{f}$ emerging from $P$ and $P'$, according to equipollence relation in $S_N$.

Hence, Levi-Civita highlights that this parallelism condition, in $V_n$, a priori depends on the path joining
$P$ with $P'$ and relying on $V_n$, being independent of it only in ordinary Euclidean spaces (which are flat).
Now, we have to specify condition \eqref{para}, by analyzing the geometric behavior of $\vec{\alpha}$ and
$\vec{\alpha}'$, supposed to be parallel between them, when $P$ moves towards $P'$ along a generic curve passing
for $P$ and $P'$, and fully relying on $V_n$. This geometrical sight will give rise to the notion of parallelism
in $V_n$.

Levi-Civita considers a generic metric on an arbitrary finite-dimensional manifold\footnote{Cf. \cite{bia2}, Ch.
XXV.} $V_n$, of the type
\begin{equation}
ds^2=\sum_{i,j=1}^n a_{ik}dx_idx_j,\label{metrica}
\end{equation}
then, he embeds $V_n$ in a Euclidean space $S_N$ with sufficiently great dimension $N\leq n(n+1)/2$, so that it
may be described by the system of equations\footnote{Cf. \cite{leci2}, eq. (1), p. 4.}
\begin{equation}
y_{\nu}=y_{\nu}(x_1,...,x_n),\qquad\nu=1,2,...,N,\label{vincoli}
\end{equation}
where the $y_\nu$ are (Cartesian) coordinates in $S_N$, while the $x_n$ are intrinsic (or Lagrangian)
coordinates on $V_n$.

Now, we observe that the functional system \eqref{vincoli} may be thought as the configuration space of a
constrained mechanical system with $n$ degrees of freedom subjected to $N$ smooth holonomic bilateral
constraints, which identify a differentiable manifold structure of dimension $n$. This is the central point of a
possible mechanical interpretation of Levi-Civita's parallelism notion: the shift from a point on $V_n$ to
another infinitesimally nearby, undergoes \eqref{vincoli} in the corresponding conceptual analogy that refers to
an holonomic mechanical system\footnote{Cf. \cite{leci}, Vol. II, Part I, Ch. V, Sect. 9, No. 63; Vol. II, Part
II, Ch. XI, Sect. 4, No. 15.} of material points with unitary mass, whose kinetic energy $T$, with respect to
\eqref{metrica}, is such that $2Tdt^2=\sum_{i,j}a_{ij}dx_idx_j$.

For simplicity, Levi-Civita considers unit vectors, hence a direction of $S_N$, arbitrarily fixed, identified by
the unit vector $\vec{\alpha}$, with direction cosines $\alpha_{\nu}$, and an auxiliary direction of $S_N$,
identified by the unit vector $\vec{f}$, with direction cosines $f_{\nu}$, $\nu=1,2,...,N$; both are supposed
emerging from an arbitrarily fixed point $P$ of $V_n$, but immersed into $S_N$. Therefore, the direction cosines
of both unit vectors $\vec{\alpha}$ and $\vec{f}$, are computed with respect to $S_N$. All that is formally
licit as $V_n$ is embedded in the ambient space $S_N$, so each direction belonging to $V_n$ also belongs to
$S_N$.

The point $P$ may be thought as a material point (with unitary mass) aimed by a certain movement along an
arbitrary smooth curve $\mathcal{C}$ lying on $V_n$, parameterized by the curvilinear (or natural) abscissa $s$
as in \eqref{metrica}, so that $\alpha_{\nu}=\alpha_\nu(s)$, $\nu=1,2,...,N$. Let $x_i=x_i(s),\ i=1,2,\dots,n$
be the intrinsic parametric equations of $\mathcal{C}$. Then, $\mathcal{C}$ may be also represented by the
parametric equations $y_{\nu}=y_{\nu}(s),\ \nu=1,2,...,N$, when it is thought embedded in $S_N$ via
\eqref{vincoli}. Indeed, since $x_i=x_i(s),\ i=1,2,...,n$, we have
\begin{equation}
 y_{\nu}=y_{\nu}(x_1(s),...,x_n(s)),\qquad\nu=1,2,...,N.\label{curva}
\end{equation}
It is evident that, in the analog constrained system of above, $\mathcal{C}$ is a trajectory in the manifold of
admissible configurations $V_n$, parameterized by time $t$ according to the parametric equations
$x_{\nu}=x_{\nu}(t),\ \nu=1,2,...,N$, with $t\in\mathbb{R}^+$.

To find a generic unit direction emerging from an arbitrary point $P$ of $\mathcal{C}$, Levi-Civita
derives\footnote{Cf. \cite{leci2}, (4), p. 5.} its parametric representation, given by \eqref{curva}, with
respect to the natural abscissa $s$
\begin{equation}
y_{\nu}'=\sum_{i=1}^n\frac{\partial y_{\nu}}{\partial x_i}x_i',\qquad\nu=1,2,...,N,\label{param}
\end{equation}
so obtaining the direction cosines with respect to $S_N$ (i.e., $y_{\nu}'$), while $x_i'$ are the direction
cosines of the same unit direction but with respect to $V_n$.

Hence, Levi-Civita considers, in a certain point $P$ of $\mathcal{C}$, an arbitrarily fixed direction
$\vec{\alpha}$ of $V_n$, whose direction cosines are $\xi^{(i)},\ i=1,2,...,n$ with respect to $V_n$, and
$\alpha_{\nu},\ \nu=1,2,...,N$ with respect to $S_N$, so that, by the {\it ansatz}
$y_{\nu}'\rightarrow\alpha_{\nu}$ (in $S_N$), $x_i'\rightarrow\xi^{(l)}$ (in $V_n$), from \eqref{param} it also
follows\footnote{Cf. \cite{leci2}, Eqs. (7), p. 6.}
\begin{equation}
\alpha_{\nu}=\sum_{l=1}^n\frac{\partial y_{\nu}}{\partial x_l}\xi^{(l)},\qquad\nu=1,2,...,N,\label{parame}
\end{equation}
which is a linear form on $\xi^{(l)}$, i.e., the direction cosines of $\vec{\alpha}$ with respect to $V_n$.

When $P$ varies along $\mathcal{C}$, imagined as aimed by a movement on $V_n$, ordinary parallelism in $S_N$
implies the equality of the angle between $\vec{\alpha}$ and an auxiliary direction $\vec{f}$ arbitrarily fixed
in $S_N$, according to Euclidean condition \eqref{para} (of synthetic geometry). Now, starting from this stance,
Levi-Civita gradually introduces an intrinsic notion of parallelism in $V_n$, considering two nearby points $P$
and $P'$ of $\mathcal{C}$, lying on $V_n$, with $P$ moving to $P'$ along $\mathcal{C}$, never leaving out $V_n$.
Therefore, in considering the arbitrarily fixed direction $\vec{f}$ of $S_N$, with direction cosines $f_{\nu}$
(in $S_N$), it follows that the cosine of the angle between $\vec{f}$ and $\vec{\alpha}$, in $S_N$, is given by
\begin{equation}
\cos\left(\widehat{\vec{f},\vec{\alpha}}\right)=\sum_{\nu=1}^N\alpha_{\nu}f_{\nu}.\label{scal}
\end{equation}

Then, Levi-Civita considers an infinitesimal variation $ds$ of the natural abscissa $s$ on $V_n$ along
$\mathcal{C}$, which implies that the cosine provided by \eqref{scal}, undergoes the following first-order
variation\footnote{In regard to this first-order variation, the arbitrariness with which $\vec{f}$ may be fixed,
implies that such an auxiliary direction, defined according to the equipollence relation in $S_N$, may be
considered as independent of $s$, at least locally in $P\in V_n$ (i.e., in $T_P(V_n)$ -- see later). Instead,
the direction $\vec{\alpha}$ a priori depends on $s$ as it varies with the movement of $P$ along $\mathcal{C}$
on $V_n$, even in a neighborhood of $P$. The equipollence relation as settled in $S_N$, locally restricted to a
neighborhood of $P\in V_n$ and defined at varying of all the curves $\mathcal{C}(\subseteq V_n)$ passing by $P$,
will led to the individuation (and, later, to the modern definition) of the so-called {\it tangent space}
$T_P(V_n)$ to $V_n$ in $P$. In \cite{leci2}, Levi-Civita shall consider $\vec{f}$ as belonging only to
$T_P(V_n)$, and not to the whole $S_N$ as in \eqref{para}, for reaching his notion of parallelism on $V_n$.}
\begin{equation}
d\cos\left(\widehat{\vec{f},\vec{\alpha}}\right)=ds\sum_{\nu=1}^N\alpha_{\nu}'(s)f_{\nu}.\label{varscal}
\end{equation}

Now, the ordinary parallelism in $S_N$ between the two directions $\vec{\alpha}(s)$ (in $P$) and
$\vec{\alpha}(s+ds)$ (in $P'$), as expressed by \eqref{para}, would require \eqref{varscal} to vanish when
$\vec{f}$ varies arbitrarily in $S_N$, so implying $\alpha_{\nu}$ to be constant or uniform. But, just at this
point, Levi-Civita introduces the key argument which will led to his notion of parallelism on $V_n$. Indeed,
since the main aim of Levi-Civita was to determine the curvature of a Riemannian manifold $V_n$, he started to
approach this problem from an intuitive viewpoint, that is to say, working out initially a geometrical setting
of the problem, then carrying on with its analytical formulation within absolute differential calculus
framework.

As has been already said above (Section 1), the usual way to determine the curvature of a Riemannian manifold
$V_n$, consists in the circuitation of a given vector around a suitable infinitesimal closed path entirely lying
on the given manifold, usually a ''parallelogrammoid'' whose sides are first-order infinitesimal geodetic
traits\footnote{This procedure is deeply argued in \cite{fok}, where, in discussing of Riemannian curvature, the
author retakes some previous considerations made by Jan A. Schouten on absolute differential calculus (cf.
\cite{schou}) and Levi-Civita's parallelism, mainly exposed taking into account the possible mechanical
interpretations most of which falling into the kinematics of rigid bodies. In this regard, we shall return on
Schouten's work in the next section.}, and drew around an arbitrary point $P$ of $V_n$. This vector should be
rotated, all around this circuit, in a parallel manner\footnote{Just according to the new definition of
parallelism as introduced by Levi-Civita in \cite{leci2}.} to itself, in such a way that, after a complete
circuitation, once reached the same point from which it departed, the possible deviation angle between initial
and final directions of such a vector in this same final and initial point, will provide an estimate of the
curvature of $V_n$.

It follows therefore the extreme importance to have a preliminary notion of parallelism on a generic Riemannian
manifold before to determine the curvature of the latter, and, to this end, as already said, Levi-Civita gave a
first geometrical sketch to this formal problem just making appeal to analytical mechanics. In this geometrical
sight of the question, Levi-Civita felt the need to consider only circuitations of a generic applied vector (as,
for instance, $\vec{\alpha}$) whose application point $P$ always relies upon the manifold $V_n$, never leaving
out it.

This circuitation, which therefore takes place exclusively upon $V_n$ and in a neighborhood of $P$, should have,
according to this Levi-Civita's geometrical framework, the main intuitive purpose, so to speak, ''to feel the
shape of $V_n$'', its distortions, cambers, deformations, and so on. And, the only intuitive way to accomplish
this end, is just the one warranting that such a circuitating vector, say $\vec{\alpha}$, remains always in
relationship with some intrinsic geometric entity characterizing $V_n$ -- i.e., $T_P(V_n)$ -- during its
circuitation, which is the result of the composition of sequential shifts along infinitesimal traits of smooth
curves on $V_n$ (usually, geodesic curves). Therefore, all that has been just said, should also hold along each
of these infinitesimal (curve's) traits.

Since a Riemannian manifold is characterized by the main property to be locally like some Euclidean space $S_N$,
then it follows that the tangent space, say\footnote{Here, we use a modern notation for the tangent vector space
to a Riemannian manifold, yet not used by Levi-Civita in his 1917 memoir, who simply speaks of ''a lying of
$S_N$ tangents in $P$ to $V_n$'' (\cite{leci2}, p. 2).} $T_P(V_n)$, at $V_n$ in some its point $P$, is just that
geometric entity characterizing better the local differentiable structure (following the well-known Hermann Weyl
work of 1913 on Riemann's surfaces) underlying any Riemannian manifold. Therefore, the Euclidean condition
\eqref{varscal} will have an intrinsic meaning related only to $V_n$ when the variability of $\vec{f}$ is
restricted from the whole $S_N$ to $T_P(V_n)$, so guaranteeing the above required reference of $\vec{\alpha}$ to
$V_n$ during its movement upon $V_n$. This intrinsic geometrical requirement, in particular, should hold too for
\eqref{varscal}.

At the same time, the variability's restriction given by $\vec{f}\in T_P(V_n)$, also guarantees that, during the
infinitesimal movement of $\vec{\alpha}$ along $\mathcal{C}$ on $V_n$, this latter ''smooths out'' $V_n$, so
''feeling'' its local curvature while $\vec{\alpha}$ moves. This is just the key geometrical intuition had by
Levi-Civita in thinking how constructively define a possible notion of parallelism on a generic Riemannian
manifold. Afterwards, this geometrical idea had to be formulated in analytical fashion, and to this end,
Levi-Civita made appeal to his wide and deep knowledge of analytical mechanics, to be precise, calling into
question the principle of virtual works in its widest and pregnant geometrical meaning.

Levi-Civita, thus, claimed that, to this end, the directions $\vec{f}$ must be exactly those compatible with the
constraints\footnote{Cf. \cite{leci2}, p. 7.} (\ref{vincoli}), once having assumed valid that pattern analogy
which considers $P$ as a material point (with unitary mass) subjected to the smooth constraints (\ref{vincoli}),
by which $\vec{f}$ must lie on $T_P(V_n)$, that is to say, $\vec{f}$ must be correlated, in this mechanical
analogy, with first-order displacements compatible with constraints (\ref{vincoli}). In such a case, while $P$
moves along $\mathcal{C}$ on $V_n$, the direction $\vec{\alpha}$ must be gradually transported always with
respect to $\vec{f}\in T_P(V_n)$, hence compatibly with the smooth constraints (\ref{vincoli}), if we wish to
define a vectorial displacement (of $\vec{\alpha}$) which must be intrinsically (correlated with or) related to
$V_n$.

At this point, still inside this geometrical framework worked out in the analytical mechanics context,
Levi-Civita, in looking at the formal aspect of \eqref{varscal}, descries a kind of physical work in $S_N$ made
by some active forces\footnote{Cf. \cite{leci0}, p. 120.}, in a certain sense corresponding to
${\alpha}_{\nu}'(s)$ (in their Cartesian components), applied to the material point $P$ (with unitary mass)
moving along $C$ with respect to the smooth constraints (\ref{vincoli}) identifying $V_n$ as an holonomic
mechanical system. So, he was legitimated to see in \eqref{varscal}, within the above mechanical analogy, the
formal expression of the principle of virtual works according to \eqref{elvlagr}, when we replace $f_{\nu}\in
T_P(V_n)$ with quantities, say $\delta y_{\nu}$, proportional to first-order virtual displacements compatible
with smooth constraints (\ref{vincoli}), that is to say, $\delta y_{\nu}$.

Therefore, with the {\it ansatz}\footnote{We cannot consider the correspondence $\vec{R}\rightarrow\vec{\alpha}$
instead of $\vec{F}\rightarrow\vec{\alpha}$, because, a priori, not always $\vec{\alpha}$ is normal to $V_n$ as
required by reactions to smooth constraints. Therefore, in applying the principle of virtual works in this
pattern analogy of Levi-Civita, we should take into account the symbolic equation of dynamics (as involving
active forces $\vec{F}$), rather than the symbolic equation of statics (as involving reactions $\vec{R}$).}
$\vec{F}\rightarrow\vec{\alpha}$, $\delta\vec{P}\rightarrow\vec{f}$, made in \eqref{elvlagr}, hence with the
further replacement of $\vec{f}=(f_{\nu})\in T_P(V_n)$ (in $S_N$) with $\delta y_{\nu}$ (in $V_n$), from $\delta
L=0$, it follows that the Euclidean parallelism condition on $V_n (\hookrightarrow S_N)$, given by the vanishing
of \eqref{varscal}, reduces to
\begin{equation}
\sum_{\nu=1}^N\alpha_{\nu}'(s)\delta y_{\nu}=0,\label{lavvirt}
\end{equation}
for any variation $\delta y_{\nu}$, that is, ''for any admissible first-order displacement compatible with the
constraints'' \eqref{vincoli}, as Levi-Civita himself said textually in \cite{leci2}, p. 7. With a suitable
mechanical interpretation of the $\alpha_{\nu}'(s)$, for instance considering them as a kind of mechanical
action in $S_N$, \eqref{lavvirt} is a formulation of the virtual work principle in $S_N$ related to the smooth
bilateral holonomic system defined by \eqref{vincoli}, hence related to analytical mechanics on a Riemannian
manifold\footnote{Cf. \cite{wr}, Ch. V; \cite{rl}, Ch. V; \cite{whi}, Ch. II; \cite{lic}, Ch. VI, Sect. I, Nos.
87-89, 92; \cite{arno}, Ch. IV; \cite{ben}, Part II, Ch. V, Sect. 6; \cite{gras}, Ch. 3, Sect. 2, No. 2.6.;
\cite{maur}, Ch. 15; \cite{fas}, Ch. 1, Sects. 1.9-12, Ch. 4; \cite{spiv2}, Part III, Ch. 12.}.

The condition \eqref{lavvirt} is however related to $S_N$. But Levi-Civita solved too this formal problem as
follows\footnote{Therefore, devoid of any historical base are all those statements for which Levi-Civita's
procedure to get parallelism, was extrinsic (to some $S_N$) and not intrinsic. Levi-Civita, in the simplest and
fastest way, got intrinsic conditions for parallelism (in $V_n$) by means of the intermediary use of an
auxiliary space $S_N$.}. To get an intrinsic form, from \eqref{vincoli}, we have\footnote{Cf. \cite{leci2},
unnumbered equation before eqs. (8), p. 7.}
\begin{equation}
\delta y_{\nu}=\sum_{k=1}^n\frac{\partial y_{\nu}}{\partial x_k}\delta x_k,\qquad\nu=1,2,...,N,
\end{equation}
with $\delta x_k$ arbitrary first-order virtual displacement on $T_P(V_n)$ (in $V_n$), so that \eqref{lavvirt}
now reduces to\,\footnote{\cite{leci2}, Eq. (8), p. 7.}
\begin{equation}
\sum_{\nu=1}^N\alpha_{\nu}'(s)\frac{\partial y_{\nu}}{\partial x_k}=0\qquad (k=1,2,...,n),\label{trasppar}
\end{equation}
which are (intermediary) formal conditions, in the intrinsic variable (or Lagrangian coordinates) $x_k$, for the
parallelism of the direction $\vec{\alpha}$ moving along $\mathcal{C}$ on $V_n$. Nevertheless, in
\eqref{trasppar} there are still parameters regarding $S_N$, so that, in order to have a full intrinsic relation
in $V_n$, it is needed to involve only parameters regarding $V_n$, without any reference to $S_N$.

To this end, we replace the direction cosines $\alpha_{\nu}(s)$ (in $S_N$) with their expression given by
\eqref{parame}, in such a way to involve only the intrinsic direction cosines $\xi^{(i)}(s)$ (in $V_n$). So,
developing related formal passages, we finally deduce\footnote{Cf. \cite{leci2}, Eq. (I$_a$), p. 8.}
\begin{equation}
\displaystyle\frac{d\xi^{(i)}}{ds}+\sum_{j,l=1}^n\Gamma^i_{jl}\frac{dx_j}{ds}\xi^{(l)}=0\qquad(i=1,2,...,n),\label{equaintr}
\end{equation}
where $\Gamma^i_{jl}$ are the Christoffel symbols of second kind with respect to the intrinsic coordinates $x_k$
(of $V_n$), defined as follows\footnote{Cf. \cite{bia1}, Ch. II.}
\begin{equation}
\displaystyle\Gamma^i_{jl}=\sum_{k=1}^na^{ik}\Big(\frac{\partial a_{kl}}{\partial x_j}+\frac{\partial
a_{jk}}{\partial x_l}-\frac{\partial a_{jl}}{\partial x_k}\Big)\qquad (i,j,l=1,2,\dots,n),\label{chris}
\end{equation}with $||a^{ik}||$ coefficient matrix of the reciprocal form of (\ref{metrica}). The \eqref{equaintr} are
the so-called (intrinsic) {\it Levi-Civita's equations of parallelism} on a Riemannian manifold $V_n$, equipped
with a generic metric of the type \eqref{metrica}.

They are first-order ordinary differential equations on the direction cosines $\xi^{(i)}$ of the arbitrary
direction $\vec{\alpha}$, emerging from $P$, transported, along a curve $\mathcal{C}$ on $V_n$, until up it
reaches the infinitesimal nearby point $P'$, from where a parallel direction $\vec{\alpha}'$ (to $\vec{\alpha}$)
emerges, with new direction cosines $\xi^{(i)}+d\xi^{(i)}$ such that (by \eqref{equaintr})
\begin{equation}
\displaystyle d\xi^{(i)}+\sum_{j,l=1}^n\Gamma^i_{jl}dx_j\xi^{(l)}=0\qquad(i=1,2,...,n).\label{equaint}
\end{equation}The \eqref{equaintr}, identify a (regular) linear system of ordinary differential equations on $\xi^{(i)}$, whose
related theorems of existence and uniqueness allow to determine a (unique) direction parallel to every other
preassigned one. It was the starting point for every possible notion of {\it connection} of differential
geometry, whose so-called ''coefficients'' are just the $\Gamma^i_{jl}$ of \eqref{equaint}. We spend a few words
on this last point, but in modern notation.

If the smooth curve $\mathcal{C}$ has parametric equation $x:[0,1]\rightarrow V_n$, then Levi-Civita's {\it
local} parallel transport along $\mathcal{C}$, as expressed by the differential forms \eqref{equaint},
establishes a (linear) isomorphism between (linear) tangent spaces $T_{x(t)}(V_n)$, $t\in[0,1]$ (if
$P\in\mathcal{C}$ is identified by $x(t)$), of the tangent bundle $T(V_n)\doteq\bigcup_{P\in V_n}T_P(V_n)$ (with
disjoint union), placed at infinitesimal nearby points of $V_n$. Instead, Levi-Civita's {\it global} parallel
transport along $\mathcal{C}$, as expressed by the ODE system \eqref{equaintr}, is the isomorphism, say
$\nabla_{\mathcal{C}}$, defined by
\begin{equation}
\displaystyle\nabla_{\mathcal{C}}:T_{x(0)}(V_n)\rightarrow T_{x(1)}(V_n)\label{lincon}
\end{equation}
through which, via \eqref{equaintr}, from the initial direction $(\xi^{(1)}(0),...,\xi^{(n)}(0))$, as initial
conditions to \eqref{equaintr}, we get the final direction $(\xi^{(1)}(1),...,\xi^{(n)}(1))$, as the
corresponding unique solution to \eqref{equaintr} by means of the well-known theorems of existence and
uniqueness for the regular system of first-order ordinary differential equations \eqref{equaintr}. So, we say
that the vector $\xi^{(0)}\in T_{x(0)}(V_n)$ is parallel (according to Levi-Civita) to the vector
$\xi^{(t)}\doteq\nabla_{\mathcal{C}}(\xi^{(0)})\in T_{x(t)}(V_n)$, along $\mathcal{C}(\subseteq V_n$), for every
$t\in]0,1]$ arbitrarily fixed. At varying of $\mathcal{C}$ in the set of all the possible smooth curves
$\mathcal{C}$ of $V_n$, from $\nabla_{\mathcal{C}}$ we may identify a so-called {\it linear connection} on
$V_n$, say $\nabla$, which generalizes the usual notion of directional derivative of ordinary Euclidean spaces,
to generic Riemannian manifolds\footnote{Cf. \cite{maur}, Part I, Ch. 1, Sect. 1.1.}.

Thus, the formal deduction of the intrinsic conditions \eqref{equaintr} (or \eqref{equaint}), characterizing
Levi-Civita's notion of parallel transport of the generic direction $\vec{\alpha}$ along an arbitrary
curve\footnote{Levi-Civita also considered (in \cite{leci2}, Sect. 7) the particular case in which $\mathcal{C}$
is a geodetic curve of $V_n$, but this has not been neither the general case taken into consideration in
\cite{leci2} in deducing the main equations \eqref{equaint} nor the initial motivation to his 1917 memoir.}
$\mathcal{C}(\subseteq V_n)$ as a function of its directional parameters $\xi_1,...,\xi_n$ with respect to
$V_n$, basically relies on the symbolic equation of dynamics \eqref{elvlagr}, which has allowed to deduce the
parallel conditions \eqref{trasppar}, whence \eqref{equaintr}. Just in this, is the power of Levi-Civita's
discovery, that is to say, having put into relation infinitesimally nearby points of a Riemannian manifold $V_n$
by means of a linear isomorphism (i.e., $\nabla_{\mathcal{C}}$) 'connecting' the related (linear) tangent spaces
at $V_n$. This had never been considered before Levi-Civita's work, which therefore became rightly a milestone
of differential geometry.

Therefore, we can say that the greatness, together the clarity and simplicity, of Levi-Civita's work, just rely
on the preliminary geometrical setting given to this question, worked out, as seen, within analytical mechanics
framework, and that always drove Levi-Civita's reasoning with those intuitions and insights that only a
geometrical preview may provide. So, analytical mechanics, with its pregnant geometrical meaning, was always a
constant guide-pattern in all those Levi-Civita's works in which such an ancient and noble doctrine could be
directly applied or simply considered at a conceptual analogy level. Besides to what has been said above, for
corroborating further this latter epistemological stance, at least in the case-study here discussed, we note
that Levi-Civita also considered\footnote{Cf. \cite{leci2}, Sect. 5.} the covariant systems
$\xi_{(i)}=\sum_{k=1}^na_{ik}\xi^{(k)}$, whose elements are said to be {\it moments}, associated to the
contravariant system $\xi^{(i)}$, hence he reformulated \eqref{equaint} in terms of $\xi_i$, so obtaining new
equations whose analytical form resembles a particular expression\footnote{Cf. \cite{leci2}, Sect. 5, Eqs.
($I_c$), p. 12; see also \cite{finpas}, Ch. X. and \cite{ago}, Vol. I.} of Lagrange's equations of dynamics on a
Riemannian manifold (Riemannian mechanics\footnote{Cf. \cite{ago}, Vol. I.}).

In conclusion, it is clear that, at least till to the definition of intrinsic parallelism on a Riemannian
manifold, the deduction of Levi-Civita's intrinsic equations of parallelism was mainly carried out within a
formal framework highly characterized by a guiding geometrical meaning coming from analytical mechanics on a
Riemannian manifold. In particular, \eqref{equaint} are deduced from \eqref{trasppar}, that is to say, the
former are nothing but a particular reformulation of the symbolic equation of dynamics for a Riemannian
manifold. In the conclusions of \cite{ir}, a deep historiographical investigation proving the constant allusion,
more or less tacit, by Levi-Civita to the principle of virtual works in working out his intrinsic equations of
parallelism, has been achieved. Only later, Levi-Civita was even more explicit in recognizing this, as, for
instance, done in treating and arguing on his notion of parallelism respectively in \cite{leci5} and
\cite{leci0}.

In this note, we have stressed this aspect of the fundamental Levi-Civita's memoir of 1917 because, from this
moment onward, many other renowned mathematicians retaken the basic ideas here exposed, to open other, fruitful
ways in mathematics, above all in differential geometry and its applications to physics. Here, we mention only
the basic work of \'{E}lie Cartan in Riemannian geometry, which started just from this Levi-Civita's memoir of
1917, above all with the acquisition of that particular method (above discussed) used to deduce the intrinsic
equations of parallelism \eqref{equaintr}. To be precise, Cartan extensively used as well analytical mechanics
concepts and methods in pursuing his work on affine connections and holonomic spaces, quoting frequently
Levi-Civita's work, taken as a guide-model together the consistent work of Gaston Darboux on the geometry of
surfaces, in whose celebrated four-volumes treatise {\it Le\c{c}ons sur la Th\'{e}orie G\'{e}n\'{e}rale des
Surfaces} of the 1880s, the fundamental elements of analytical mechanics on a Riemannian manifold are
exposed\footnote{Cf. \cite{dar}, Tome II, Livre V, Chapitre VIII. Soon after Darboux's treatise, also the
celebrated many volumes Paul E. Appell's {\it Trait\'{e} de m\'{e}canique rationnelle} of the 1890s, deals with
the first elements of the basic analytical mechanics from a Riemannian viewpoint, taking the legacy of the
previous works made by Rudolf Lipschitz, Joseph Liouville, Joseph Bertrand, Edouard J.B. Goursat and others, on
the analytical foundations of rational mechanics (cf. \cite{lut}). One of the first textbooks devoted to a
comprehensive treatment of analytical mechanics, with first explicit recalls to Riemannian geometry, was
\cite{whi} (see, in particular, Ch. II). In any case, in regard to the history of analytical mechanics on
Riemannian manifold contextual with the historical question treated in the present paper, it might be useful,
from an historiographical standpoint, to look at the few references consulted by Levi-Civita and Amaldi in
drawing up their treatise \cite{leci} (see, in particular, \cite{leci}, Vol. I, p. VI).}.

\section{Further historical remarks}
Besides to what has been exposed in \cite{ir}, in this last section we wish to outline further historical
remarks on the celebrated work of Levi-Civita on parallel transport, \cite{leci2}. To be precise, we would like
to point out, another time, the extreme originality and uniqueness of the method followed by Levi-Civita in
pursuing his scopes in \cite{leci2}. The customary use of analytical mechanics methods in his mathematical
studies, was really original and an usual working praxis which led him to reach the highest results, as the one
treated above. Many influential historical sources cite other possible analogous case-studies, just regarding
the crucial question concerning parallelism in a generic Riemannian manifold, but none of these may be compared
with the greatness and originality of Levi-Civita's insight, formally carried on with an impeccable stylistic
elegance\footnote{In this regard, see also the opinion expressed in \cite{juv}, p. 73.} and with an amazing
simplicity of method. We briefly discuss some of these sources, which make manifest reference to \cite{leci2}.

Marcel Berger stated\footnote{Cf. \cite{ber}, Ch. XV, Sect. 15.3.}, for instance, that Levi-Civita is placeable
into the so-called {\it golden triangle} of Riemannian geometry, whose vertices are: the {\it curvature} (via
the {\it connection}), the {\it parallel transport} and the {\it absolute differential calculus}. This triangle
was first understood by Ricci and Levi-Civita at the end of 19th century. Berger says also that the basic {\it
lemma}, which is the key to everything, is just the existence and the uniqueness of a canonical connection,
called the {\it Levi-Civita's connection}, on any Riemannian manifold.

According to Berger, it is important to realize that this lemma is a ''miracle'': many people have tried to
understand it, with more or less sophisticated concepts, but we consider that it remains a miracle. This might
perhaps explain why there have been many variations in the historical interpretations of this fundamental notion
of mathematics. In this note, we have shed some light on this ''miracle'', trying to clarify historically that
has been the right place in which such a ''miracle'' happened, finding in the analytical mechanics framework, as
seen above, the right context in which it occurred as such.

Doing reference to analytical mechanics has been a typical and fruitful praxis of Levi-Civita's working. Indeed,
besides to what has been said above on \cite{leci2}, another historiographical prove of the predominance of
geometrical sight in formulating and approaching a general formal problem by Levi-Civita, and arising above all
from the analytical mechanics framework, may be found in a work immediately following \cite{leci2} but closely
related to the same research program centred on formal aspects of general relativity, namely \cite{leci4}, where
Levi-Civita gave an extremely interesting and clear interpretation of the Einstein's field equations of general
relativity consisting in giving a generalized form to D'Alembert principle of analytical mechanics in the
relativistic context, got by means of a geometrical sight, ingeniously had by Levi-Civita, of certain
geometrical relations involving first-order covariant derivatives of Riemann's symbols, due to Luigi Bianchi
and, therefore, said to be {\it Bianchi's identities}. The final result is now a formally correct
expression\footnote{Cf. \cite{leci2}, Sect. 8, Eqs. (10'), p. 55. Among other, in this paper, Levi-Civita gave,
for the first time, the real, formally correct expression of Einstein's field equations which were initially set
up, by Einstein himself, in a not properly correct form from the analytical standpoint. To this formal lack,
Levi-Civita remedied, still again, through his powerful and ingenious method consisting in doing reference,
whenever possible, to analytical mechanics, trying to lay out in its framework the given formal problem, to be
better and easier approached and solved, with the further possibility to have an enlightening physical
interpretation (cf. \cite{pal2}, Sect. 6). However, right criticisms to the formal correctness of first forms of
Einstein's field equations were moved by Levi-Civita to Einstein since the first months of 1915, with a thick
correspondence between them (cf. \cite{buc}, Vol. 8), which allowed Einstein to give finally the right and
formally correct expression to his celebrated field equations, around the end of 1915, almost in the same period
in which Hilbert wrote his known works on the related variational aspects of the question, following the
analytical mechanics pattern of Hamilton's variational principle (cf. \cite{pal2}).} of Einstein's field
equations in terms of an extended D'Alembert principle of analytical mechanics to general relativity, so
reaching new and proper physical interpretations of the field equations themselves.

Furthermore, in the authoritative Klein's {\it Encyklop\"{a}die der mathematischen Wissenschaften mit
Einschlu\ss ihrer Anwendungen}\footnote{Cf. \cite{enc}, Band 3, Teilen 3, III.D.11-B.II.18.}, which is one of
the widest and richest bibliographical sources of the time, there are some other interesting references usually
not quoted elsewhere, regarding Levi-Civita's parallel transport and its possible links with mathematical
physics. Among these, a work of Adrian D. Fokker, namely {\cite{fok}, which is even quoted, in this
encyclopedia, as the only one, of that time, to have given a geometrical and mechanical interpretation of
Levi-Civita's parallelism.

In this Fokker's paper, however, there is an interesting exposition of the usual geometrical method to calculate
Riemann curvature by means of the circuitation of a vector around an infinitesimal closed path whose sides are
infinitesimal geodesic traits, according to some ideas of Schouten (on the so-called {\it geodesic
displacement}), where frequent recalls to kinematics of rigid body are done in regard to certain {\it geodesic
displacements} considered by the author. The principle of the method, nevertheless, which surely makes reference
often to rigid mechanics notions, seems to be very close to Cartan's method of moving frame (which appears to
anticipate this last in many of its respects), but does not have to do, in no way, with Levi-Civita's one.

Likewise to the case of Fokker's paper, in this enciclopedia\footnote{Cf. \cite{enc}, Band 3, Teilen 3,
III.D.11-B.II.18, p. 131.} is also quoted the well-known William Thompson (Lord Kelvin) and Peter G. Tait
two-volumes {\it Treatise on Natural Philosophy} of 1879, in regard to first notions of parallelism on a
surface, built up by means of kinematical methods\footnote{See also \cite{dar}, Tome II, Livre V, Ch. VII, for
the analogy of the method.}. Nevertheless, what is exposed in this popular treatise on mathematical physics
about this argument\footnote{Moreover, retaken later by \cite{col}.}, does not have any conceptual link with the
original method of Levi-Civita which, as has been already said above, makes use, originally and fruitfully, of
higher analytical mechanics methods and concepts.

Finally, also Vladimir I. Arnold, after having rightly recalled\footnote{Cf. \cite{arn}, Ch. 6, Sect. 6.4.3, pp.
331-32.} that, in Riemannian geometry as well as therefore in general relativity theory and in gauge theory of
field theory, a fundamental role is played by Levi-Civita's connection (defined through the coefficients
$\Gamma^i_{jl}$ of $\eqref{equaint}$), which defines parallel transport of vectors along a manifold with a
generic Riemannian metric, stated nevertheless that, the most physically natural definition of this (quite
non-obvious) vector transport on a Riemannian manifold was first provided by Johann Radon in a work\footnote{Cf.
\cite{kle}, Part III, Ch. II. Strangely enough, the only, official source which reports this Radon's work, is
\cite{kle}, Part III, Ch. II, which is just devoted to the exposition of the mechanical interpretation provided
by Radon in 1918 to Levi-Civita's parallelism.} of 1918 on the early theory of adiabatic invariants, which
seems, according to Arnold, having been motivated just from Levi-Civita's memoir of 1917.

To be precise, Radon imagined a conceptual physical experiment\footnote{Levi-Civita, instead, follows, from an
epistemological viewpoint, a pattern analogy settled between mathematical physics and geometry.} ({\it
Gedankenexperiment}) in which is placed, at an arbitrary point of a given Riemannian manifold, some oscillatory
system as, for example, the one got suspending a Foucault's pendulum over this point, or considering, in the
tangent space at this point, a Hooke's elastic system with potential energy proportional to the square of the
distance from the original point. Then, under suitable initial conditions, this oscillatory system is conceived
to perform a so-called {\it eigenoscillation} in the direction defined by some generic vector of the tangent
space. Hence, if this oscillatory system is being transported, slowly and smoothly, along some path lying on the
given manifold, then it follows, from adiabatic theory, that the oscillation will remain (in the adiabatic
approximation) an eigenoscillation. Furthermore, its direction (i.e., the polarization) will rotate somehow
during the motion of the point along the path, this rotation -- which proves to be an orthogonal linear
transformation of the initial tangent space into the terminal one -- Radon observed to be just due to
Levi-Civita's parallel transport (or connection) in dependence on the curvature of the manifold.

It is historically interesting to note that Radon's theory on adiabatic invariants was not understood by
geometers of the time, mainly because they were not familiar with these latter, so that it was unfairly
forgotten. Arnold, however, above defined this Radon's interpretation as the first most physically natural
definition of Levi-Civita's parallel transport, which sounds us a quite strange, seen the great and deep
expertise in mathematical physics, even more in mechanics, owned by Arnold, who, maybe, has never read the
original Levi-Civita's memoir that, strangely, was never translated in any foreign language. Indeed, everyone
knows the minimal requisites of analytical mechanics, in reading the original Levi-Civita's memoir, immediately
recognizes the yet tacit use of founding principles of analytical mechanics. Nonetheless, Radon's interpretation
is, surely, a very clever one, although quite cumbersome and, however, relegated to a mere thought experiment.

There is, furthermore, another historical aspect of the history of Levi-Civita's parallel transport, concerning
a question of discovery's priority, that deserves here to be clarified briefly. Precisely, Schouten
claimed\footnote{Cf. \cite{enc}, Band 3, Teilen 3, III.D.11-B.II.18, p. 131; \cite{str}, \cite{st}, \cite{stu}.}
the priority in discovering a notion of parallelism in a Riemannian manifold, called {\it geodesic displacement}
and exposed in \cite{schou}, which should include, according to him, the Levi-Civita's one\footnote{Cf.
\cite{schou}, footnote a), p. 46.}. He stated that this his paper had already been drawn up since 1915, but that
it was yet published (with an unusual and strange delay) only in 1919 in the proceedings of the Academy of
Sciences of Amsterdam, inasmuch officially classified as ''Verschenen Februari 1919'', i.e., published on
February, 1919. In this regard, some oral testimonies say that Luitzen J. Brouwer, who also contributed
initially to the subject and was a colleague of Schouten, opposed to the revindication of the latter, giving
priority to Levi-Civita.

Anyway, what is really important in historiography, no matter any other testimony, is the certainty and
officiality\footnote{Like, for example, the registration of a possible preprint in some lawcourt or in a
suitable inventory deposited in some academic department office or library. On the other hand, most of the
academic journals which host publications of various type, often report the so-called {\it publication history},
i.e., date of submission or receipt of the paper in its first version, dates of further revisions, date of
acceptance, and so on. All that, allows, for instance, to have objective historiographical data. This has not
been possible for Schouten's paper, to confirm the priority claimed by him.} in dating the related historical
sources used as such, and, until up now, there are no reliable official historical sources, except little
reliable oral witnesses, that objectively corroborate what Schouten claimed. So, until proven otherwise, his
work remains however dated\footnote{From a direct inspection of the original issue where Schouten's work was
published, it is turned out that, in the table of contents ({\it Inhoud}) of this issue (in which seven
contributions are listed in chronological order -- the Schouten's one, is the 6th -- and everyone with its own
internal page enumeration), is reported the publication date of February, 1919 ({\it Verschenen Februari 1919}),
while in the title-page of the contribution, is instead reported the year 1918. So, there is no certainty
neither on this historical datum. Nevertheless, from an historiographical standpoint, we should consider the
date of 1919, the only one to have been explicitly quoted as such, i.e., as a ''publication date'' ({\it
Verschenen Februari 1919}). Instead, just at the end of \cite{leci2}, there is the date of final writing of the
paper by Levi-Civita, that is, November, 1916, while at the beginning of \cite{leci2}, there is the date in
which this memoir was presented, the 24th of December 1916, at the periodic monthly sessions of the {\it Circolo
Matematico di Palermo}, before it were being sent to the related {\it Rendiconti del Circolo Matematico di
Palermo} to be officially published. In any way, even in this further case, there is no right comparison with
the great intuitive power and the wide prospective amplitude opened by Levi-Civita's method.} to 1919. Likewise,
some sources quote the work of Gerhard Hessenberg \cite{hes} as contemporary to that of Levi-Civita, in
introducing a notion of parallelism on a Riemannian manifold, following a vectorial method. In any case, this
last work was realized fully within Ricci's and Levi-Civita's absolute differential calculus framework, without
any useful geometrical consideration or insight, differently from Levi-Civita's one, whose greatness relies just
in its deep and immediate geometrical intuition of the crucial idea underlying the fundamental notion of
connection.

Finally, according to a remark due to Jacques Hadamard\footnote{Mentioned in \cite{juv}, Ch. VI, p. 73, but
without having further, more detailed, bibliographical indications neither to this Hadamard's affirmation nor to
the related Darboux's work (cf. \cite{dar}, Tome II, Livre V, and, above all, \cite{dar1}, Livre II, Ch. II).},
as early Darboux, in dealing with geodesic curvature of an ordinary surface computed by means of his method of
moving frame (Darboux's {\it repère mobile}), had already reached a similar notion of parallelism, yet without
having been rightly recognized as such. But, even if surely Darboux's oeuvre established a deep, wide and
interesting intertwinement between geometry and mechanics\footnote{See, above all, \cite{dar}, Tome II, Livre V,
Chapitres VI-VIII.}, in \cite{dar} he restricted the study only to ordinary surfaces in Euclidean spaces, while
in \cite{dar1}, he effectively considered\footnote{See \cite{dar1}, Livre II, Chapitre II, NN. 115-16, where he
decomposes the infinitesimal quadratic element $ds^2$ in the sum of three independent infinitesimal components
related to rotations and translations of the repère mobile. A similar method, but extended to the
$n$-dimensional case, will be then retaken by Anita Carpanese in \cite{car} to compute curvature of a generic
Riemannian manifold, again starting from Levi-Civita's result.} the case of the parallel displacement, upon a
three-dimensional Riemannian manifold, of a {\it trièdre mobile} moving along a curve lying on a Riemannian
manifold, but with methods, approaches and results which didn't have that immediate and powerful geometric
intuition owned by Levi-Civita's work, which distinguishes for its higher contextual coherence, conceptual
clearness and great potentiality, together methodological originality and generality, as those copious
achievements, immediately later pursued in differential geometry and theoretical physics, manifestly proved.
However, the right and large tribute to Darboux's work, just in this context, will come from the next,
pioneering work of Cartan on Riemannian geometry, which yet started from Levi-Civita's memoir, following that
fruitful method based on a preliminary intuitive sight (mainly, of geometric nature) which always accompanied
and led Levi-Civita's formal thought.

\section{Conclusions}
In short, we may say that, after what has been said so far, as well as after having consulted most of the main
modern and past literature either in differential geometry and its applications to physics\footnote{Here, we
have quoted only those references which have been strictly essential in drawing up this note.}, above all to
general relativity, the conclusion of this our further historical investigation (besides \cite{ir}), is that
nobody has highlighted, even fleetingly, the crucial and ingenious role played by the fundamental analytical
mechanics principles in Levi-Civita's construction, notwithstanding their {\it manifest} presence in his works.
The lucky establishment of that conceptual pattern analogy between geometry and analytical mechanics, led
Levi-Civita to work out a preliminary framework in which at first lay out the original geometrical question,
then approach and solve it, achieving a so rich harvest of fruitful results that conferred an unusual success to
this method intertwining geometry and analytical mechanics.

However, from a methodological standpoint, this reflects, as Ugo Amaldi said in the biographical introduction to
\cite{leci3}, that traditional and usual methodological praxis of modern Italian mathematical school (between
the end of 19th century, and the beginning of 20th century), fundamentally based on a preliminary geometric
sight of any possible formal question before this is being treated in a pure analytical way. In the case-study
here examined, this way turned out to be extremely fruitful in mathematics, as the next work of \'{E}lie Cartan
masterfully witnessed\footnote{Cf. \cite{lau}, Ch. 2, Sect. 1.}, as well as in theoretical physics, as the next
results in field theory (above all, in general relativity) testified too. This reasoning's fashion was, almost
always, presents in all those Levi-Civita's works where such a method could run: we have only considered here,
in a detailed manner, \cite{leci2}, as well as, fleetingly, \cite{leci4}, as emblematic historical events.

As in most of his works, Levi-Civita was able not only to give a first, clarifying geometrical setting to every
formal question had to be treated, but also in using every possible tools of mathematical physics to approach
it, above all higher mechanics, maybe for its strong geometrical intuition. This was, as already said, a
powerful and successful method which featured almost all the intellectual work of Levi-Civita, as well as other
scholars, like Hadamard. Indeed, as pointed out in \cite{fg}, both Levi-Civita and Hadamard, for instance in
dealing with formal problems concerning wave propagations and their discontinuities to be treated by means of
systems of hyperbolic partial differential equations, always put before these suitable physical arguments thanks
to which outline easier the needful formal framework in which to lay out these questions to be then analytically
handled.

To testify the related methodological importance and power, as we have seen above in discussing briefly also
\cite{leci4}, this typical method of Levi-Civita was, for instance, so crucial in helping Einstein to reach
finally the definitive, correct analytical form of his celebrated field equations, which initially were affected
by some formal lacks. This was possible after having laid out the question concerning the analytical setting of
the equations of gravitational field, within an appropriate analytical mechanics framework which allowed easy
Levi-Civita to interpret these equations as a suitable extension of D'Alembert principle to relativity context,
as well as to get new physical interpretations of them\footnote{Cf. \cite{pal2}, Sect. 6.} which, with a
formidable intuition, Levi-Civita masterly used to guide his analytical investigation of gravitational field
equations, so being able to suggest to Einstein what pathway follows to reach their correct, formal expression.

The main aim of this note has been therefore to put in evidence this historical-epistemological aspect of
Levi-Civita's method in approaching mathematical problems, mainly discussing an historical case-study (namely,
\cite{leci2}) whose idea there contained turned out to be so crucial for the next development of
mathematics\footnote{Cf. \cite{daya}.} and theoretical physics, i.e., the one related to the discovery of
parallelism in a generic Riemannian manifold. As regard then the latter context, seen the wideness of the
related historical literature already existent, we think that enough is to recall only Hermann Weyl's (cf.
\cite{wey}) and \'{E}lie Cartan's (cf. \cite{cart}) works on pure differential geometry and its applications to
theoretical physics, to witness the remarkable prominence of the next developments of Levi-Civita's idea. On the
other hand, right lately, this notion has been also retaken by the current research in applied engineering, to
show the extent relevance of the topic, still now: in this regard, here we quote only two recent, interesting
works of structural mechanics applied to engineering sciences, \cite{cas} and \cite{pr}, just centred on
Levi-Civita's notion of parallel transport.

\subsubsection*{Acknowledgements} This paper springs out of an history of mathematics seminar held at the Department
of Mathematics of the University of Bologna, on July, 7, 2016, organized by Professors Sandro Graffi and
Giovanni Dore of {\it Alma Mater Studiorum}, who therefore thank so much.

\end{document}